\documentclass[aps,twocolumn,showpacs,10pt,superscriptaddress,preprintnumbers,nofootinbib]{revtex4-1}
\usepackage[colorlinks,citecolor=blue]{hyperref}
\usepackage{enumerate}
\usepackage{xcolor,graphicx}
\usepackage{latexsym,cancel,amssymb,amsmath,verbatim,mathrsfs}
\usepackage{mathtools}
\usepackage{physics,ulem}
\usepackage{subfigure}

\newcommand{\bea}{\begin{eqnarray}}
\newcommand{\beq}{\begin{equation}}
\newcommand{\eea}{\end{eqnarray}}
\newcommand{\eeq}{\end{equation}}

\allowdisplaybreaks[4]

\begin{document}

\preprint{
	{\vbox {			
		\hbox{\bf MSUHEP-23-022}
}}}
\vspace*{0.2cm}

\title{Probing the $Zb\bar{b}$ coupling at the $Z$-pole of future lepton colliders}

\author{Bin Yan}
\email{yanbin@ihep.ac.cn}
\affiliation{Institute of High Energy Physics, Chinese Academy of Sciences, Beijing 100049, China}

\author{C.-P. Yuan}
\email{yuan@pa.msu.edu}
\affiliation{Department of Physics and Astronomy,
Michigan State University, East Lansing, MI 48824, USA}

\author{Shu-Run Yuan}
\email{sryuan@stu.pku.edu.cn}
\affiliation{School of Physics, Peking University, Beijing 100871, China}

\begin{abstract}
The determination of the $Zb\bar{b}$ coupling in experiments has been a long-standing challenge, as the limited precision of off $Z$-pole measurements at the LEP has resulted in two degenerate solutions remained to be resolved.
In this paper, we propose a novel method to probe the $Zb\bar{b}$ coupling by measuring the forward-backward asymmetry of the bottom quark, $A^{b}_\text{{FB}}$, in the $b\bar{b}$ system of the $e^+ e^- \to b \bar{b} \gamma$ and/or $e^+ e^- \to b \bar{b} g$ processes at the $Z$-pole of future lepton colliders. The additional hard photon or light jet radiation can mimic the energy scanning of the $e^+e^-\to b\bar{b}$ process, and the $A_{\rm FB}^b$ distribution 
from the $\gamma$-$Z$ interference process is linearly sensitive to the $Zb\bar{b}$ coupling.
By combining the expected measurements of $R^0_b$ and $A^{0,b}_\text{{FB}}$ at the $Z$-pole at the CEPC, the $A^{b}_\text{{FB}}$ distributions can break the degeneracy observed at the LEP, leading to a unique determination of the $Zb\bar{b}$ coupling through $Z$-pole running alone.

\end{abstract}

\maketitle

\section{Introduction}
The forward-backward asymmetry of bottom quark ($A^{b}_\text{FB}$) at the $Z$-pole ($A^{0,b}_\text{FB}$) from the Large Electron-Positron collider (LEP)
exhibits a long-standing discrepancy with the Standard Model (SM) prediction, with a significance about 2.1$\sigma$~\cite{ParticleDataGroup:2020ssz}. 
This anomaly can potentially be explained by intriguing new physics (NP) models that involve a significant modification of the right-handed $Zb\bar{b}$ coupling while maintaining a SM-like left-handed $Zb\bar{b}$ coupling~\cite{Choudhury:2001hs, Agashe:2006at, Liu:2017xmc, Crivellin:2020oup}. Such modifications could arise from an approximate custodial symmetry inherent in the underlying theory~\cite{Agashe:2006at}. Additionally, the fact that the left-handed bottom quark and top quark belong to the same electroweak doublet implies that any deviation in the $Zb\bar{b}$ coupling is inevitably connected to top quark interactions~\cite{Agashe:2006at,Peccei:1990uv, Haber:1999zh,Cao:2015doa,Cao:2015qta,Cao:2020npb}. Thus, precise measurements of the $Zb\bar{b}$ coupling are crucial not only for testing NP models involving the bottom quark but also for probing the properties of the top quark.

However, the $Zb\bar{b}$ coupling can not be determined uniquely due to the quadratic dependence on this coupling of observables such as $A_{\rm FB}^{0,b}$, $R^0_b$ (the ratio of the $Z \to b \bar{b}$ partial decay width to the inclusive hadronic decay width at the LEP), and $A_b$ (the left-right forward-backward asymmetry at the Stanford Linear Collider (SLC)) at the $Z$-pole~\cite{ALEPH:2005ab}. To break this degeneracy, the measurement of $A_{\rm FB}^b$ in the off-$Z$ pole region becomes crucial as it allows for discrimination among the four degenerate solutions by taking into account the important $\gamma$-$Z$ interference effects, which are linearly dependent on the $Zb\bar{b}$ coupling~\cite{Choudhury:2001hs}. Unfortunately, due to limitations in the statistics of off-$Z$ pole data from LEP, only the degeneracy of the left-handed component coupling $Z b_L \bar{b}_L$ has been broken, while the degeneracy of the right-handed component $Z b_R \bar{b}_R$ coupling remains unresolved.

Recent studies have proposed various approaches to address this degeneracy and further investigate the anomalous $Zb\bar{b}$ couplings at the Large Hadron Collider (LHC) and future colliders, with the aim of confirming or ruling out the discrepancy between the $A^{0,b}_\text{FB}$ measurement and the SM prediction at the $Z$-pole. For example, the axial-vector component of the $Zb\bar{b}$ coupling can be determined through precision measurements of the $gg \to Zh$ production cross section and exclusive $Z$-boson rare decays $Z \to \Upsilon(ns)+\gamma$ at the LHC and high-luminosity LHC (HL-LHC)~\cite{Yan:2021veo,Dong:2022ayy}. On the other hand, the vector component of the coupling can be constrained effectively by the electron single-spin asymmetry in neutral current deeply inelastic scattering (DIS) process at HERA and the upcoming Electron-Ion Collider (EIC)~\cite{Yan:2021htf}. Additionally, the application of additional jet charge information in DIS process at the EIC can further improve the measurement of the $Zb\bar{b}$ coupling through the jet charge weighted single-spin asymmetry~\cite{Li:2021uww}. Furthermore, it has been suggested in Ref.~\cite{Bishara:2023qhe} that the degeneracy of the $Zb_R\bar{b}_R$ coupling could also be broken through the measurement of the charge asymmetry of the final state $b$ and $\bar{b}$ jets in the associated production of $Z$ boson and two $b$-jets at the HL-LHC.

At the same time, future lepton colliders such as the Circular Electron-Position Collider (CEPC)~\cite{CEPCStudyGroup:2018ghi}, International Linear Collider (ILC)~\cite{ILC:2013jhg}, and the Future Circular Collider (FCC-ee)~\cite{FCC:2018evy} offer excellent opportunities for further investigation of the $Zb\bar{b}$ coupling. With significant improvements in the statistical uncertainties of $Z$-pole measurements, the discrepancy of the $A_{\rm FB}^{0,b}$ at the LEP could be directly resolved at these future lepton colliders. Additionally, the degeneracy of the $Zb_R\bar{b}_R$ coupling could also be broken by measuring $A_{\rm FB}^{b}$ around $\sqrt{s}=240\sim 250~{\rm GeV}$ at these colliders, which is the primary collider energy of the next generation lepton collider for studying the Higgs properties~\cite{Gori:2015nqa}.

\begin{figure}
\subfigure[]{
    \centering
    \includegraphics[width=0.13\textwidth]{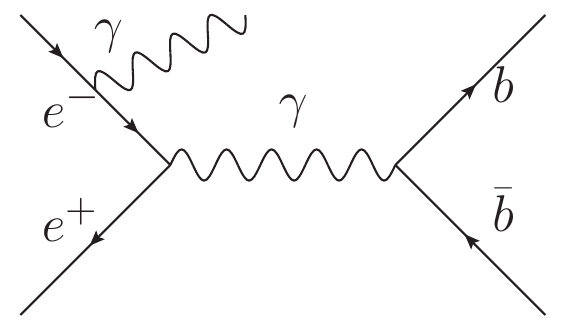}
    \label{ga_gaISR}
}
\subfigure[]{
    \centering
    \includegraphics[width=0.13\textwidth]{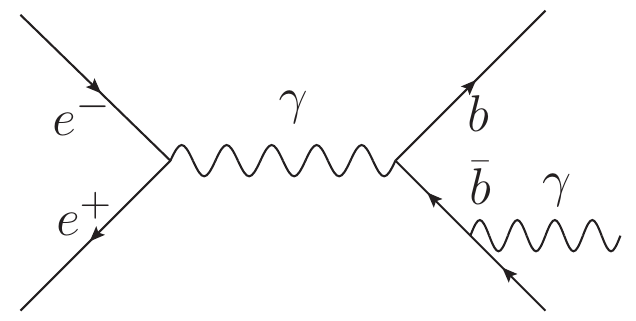}
    \label{ga_gaFSR}
}
\subfigure[]{
    \centering
    \includegraphics[width=0.13\textwidth]{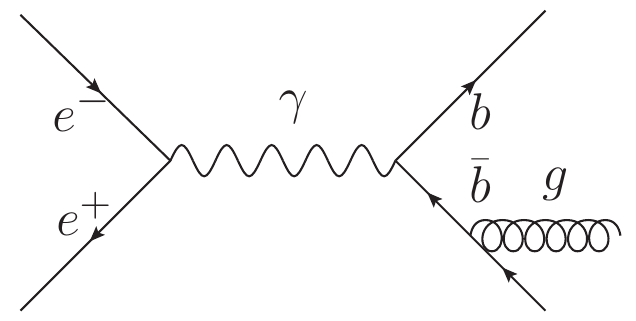}
    \label{ga_glFSR}
}

\subfigure[]{
    \centering
    \includegraphics[width=0.13\textwidth]{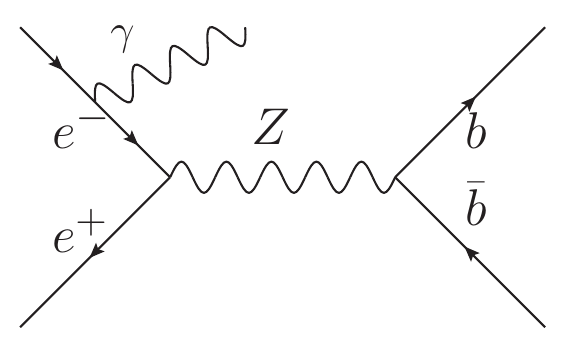}
    \label{z_gaISR}
}
\subfigure[]{
    \centering
    \includegraphics[width=0.13\textwidth]{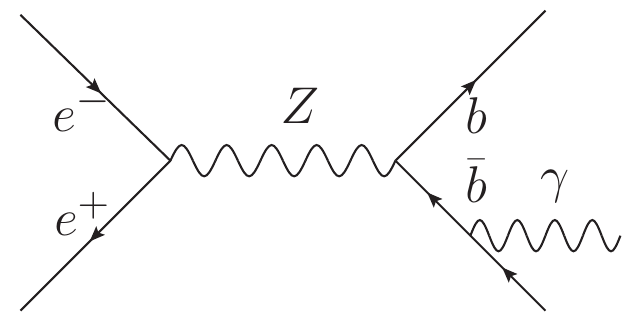}
    \label{z_gaFSR}
}
\subfigure[]{
    \centering
    \includegraphics[width=0.12\textwidth]{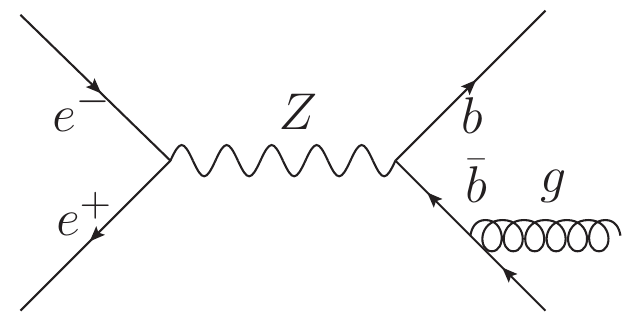}
    \label{z_glFSR}
}
\caption{
  Illustrative Feynman diagrams for the $e^+ e^- \to b \bar{b} \gamma$ and $e^+ e^- \to b \bar{b} g$ productions at the future lepton colliders.
}
\label{fig:feyn_diag}
\end{figure}

In this paper, we suggest studying the $e^+ e^- \to b \bar{b} \gamma$ and/or $e^+ e^- \to b \bar{b} g$ processes, as shown in Fig.~\ref{fig:feyn_diag}, to break the degeneracy of the $Zb_R\bar{b}_R$ coupling at the future lepton collider at the $Z$-pole.
By allowing for the emission of one hard photon or light jet, the invariant mass of the $b\bar{b}$ system, denoted as $m_{b\bar{b}}$, can be shifted away from the $Z$-pole, mimicking the energy scanning of the $e^+ e^- \to b \bar{b}$ process. Consequently, one can evaluate the forward-backward asymmetry of the bottom quark in the center-of-mass (c.m.) frame of the $b\bar{b}$ system, and the contribution from the $\gamma$-$Z$ interference is linearly sensitive to the $Zb\bar{b}$ coupling.
Due to the large amount of data that will be collected around the $Z$-pole at the future lepton collider, we demonstrate that by combining the $A^b_\text{FB}$ distributions from the $b\bar{b}\gamma/g$ processes with the $R^0_b$ and $A^{0,b}_\text{FB}$ measurements at the $Z$-pole, it is possible to uniquely determine the $Zb\bar{b}$ coupling. Therefore, our results provide important complementary information to other methods in the literature for probing the $Zb\bar{b}$ interaction.

\section{Bottom Quark Forward-Backward Asymmetry \label{sec:asymmetry}}
In this section, we discuss the bottom quark forward-backward asymmetry as a function of $m_{b\bar{b}}$ at the parton-level through $e^+ e^- \to b \bar{b} \gamma$ and $e^+ e^- \to b \bar{b} g$ processes at the $Z$-pole. In the $e^+ e^- \to b \bar{b} \gamma$ process, the photon can arise from both initial state radiation (ISR) and final state radiation (FSR), while in the $e^+ e^- \to b \bar{b} g$ channel, gluon radiation is only available from FSR; see Fig.~\ref{fig:feyn_diag}. 
To illustrate the impact of the $Zb\bar{b}$ coupling on $A_{\rm FB}^b$ as a function of $m_{b\bar{b}}$ at the $Z$-pole, we will use the $e^+ e^- \to b \bar{b}\gamma$ process as an example. The $Zb\bar{b}$ effective Lagrangian can be parametrized as
\beq
\mathcal{L}_{eff}=\frac{g_W}{c_W}Z_{\mu}\left(\kappa_L g_L \bar{b}_L \gamma^{\mu}b_L +\kappa_R g_R \bar{b}_R \gamma^{\mu}b_R\right)\,,
\eeq
where $g_W$ is the $SU(2)_L$ gauge coupling, and $g_L=-1/2+s^2_W/3$ and $g_R=s^2_W/3$ are the left and right-handed components of $Zb\bar{b}$ coupling in the SM, respectively. Here, $s_W^2\equiv \sin\theta_W^2$, with $\theta_W$ being the weak mixing angle.
$\kappa_{L,R}$ are effective parameters which encode possible NP effects in $Zb\bar{b}$ interaction and $\kappa_{L,R}=1$ for the SM. 

\begin{figure}[tb]
    \centering
    \includegraphics[scale=0.65]{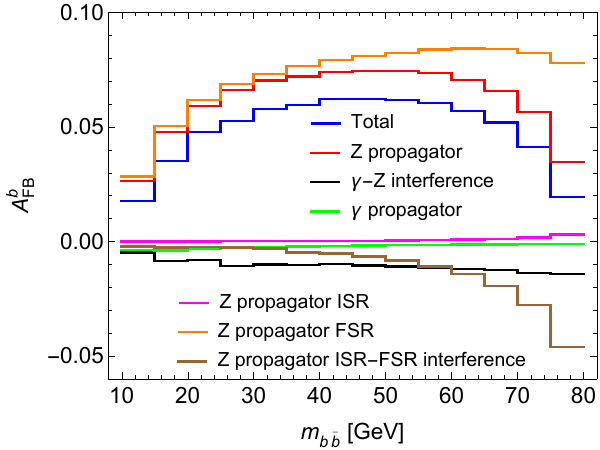}
\caption{
  The $A^b_\text{FB}$ distribution of the $e^+ e^- \to b \bar{b} \gamma$ production at $Z$-pole in the SM.
  The ``$Z$ propagator ISR", ``$Z$ propagator FSR" and ``$Z$ propagator ISR-FSR interference" correspond to Fig.~\ref{z_gaISR}, \ref{z_gaFSR} and their interference, while ``$Z$ propagator" (Fig.~\ref{z_gaISR}+~\ref{z_gaFSR}), ``$\gamma$ propagator" (Fig.~\ref{ga_gaISR}+~\ref{ga_gaFSR}) and ``$\gamma$-$Z$ propagator" are the contributions from the mediator $Z$ boson, photon and their interference. 
}
\label{fig:Afb_part}
\end{figure}

Figure~\ref{fig:Afb_part} presents the $A_{\rm FB}^b$ distributions as a function of $m_{b\bar{b}}$ for different parts of the $e^+e^-\to b\bar{b}\gamma$ process in the SM. We impose a requirement of $p_T^\gamma>10~{\rm GeV}$ for the transverse momentum of the photon to enhance the contribution of the $\gamma$-$Z$ interference. From the orange line of Fig.~\ref{fig:Afb_part}, it is evident that the dominant contribution to $A_{\rm FB}^b$ comes from the on-shell $Z$ production with a hard photon radiation from the final states (see Fig.~\ref{z_gaFSR}, labelled as ``$Z$ propagator FSR" in Fig.~\ref{fig:Afb_part}). The value of $A_{\rm FB}^b$ from this channel increases with $m_{b\bar{b}}$ until $m_{b\bar{b}}\sim 30~\rm{GeV}$, after which it flattens out. The behavior at small $m_{b\bar{b}}$ is due to the non-negligible effect of bottom quark mass, which flips the helicity of the bottom quark.  As a result, the production rate is proportional to $m_b\sin\theta$, where $\theta$ is the polar angle of the bottom quark in the $b\bar{b}$ system, so that $A_{\rm FB}^b$ becomes smaller as $m_{b\bar{b}}$ decreases. The interference effects between ISR and FSR with the mediator of the $Z$ boson (brown line, labelled as ``$Z$ propagator ISR-FSR interference" in Fig.~\ref{fig:Afb_part}; the interference between Fig.~\ref{z_gaISR} and Fig.~\ref{z_gaFSR}) also contribute significantly to $A_{\rm FB}^b$, showing a stronger dependence on $m_{b\bar{b}}$. This is because the $Z$-boson propagator in the ISR process is $1/(m_{b\bar{b}}-m_Z^2)$, cf.  Fig.~\ref{z_gaISR}. It is important to note that the $A_{\rm FB}^b$ from this interference is not a result of the parity violation of the $Zb\bar{b}$ interaction. The size of $A_{\rm FB}^b$ depends on the combination of the couplings $g_V^2+g_A^2$, where $g_V=g_L+g_R$ and $g_A=g_L-g_R$, and can be understood from the charge conjugation transformation for the final states~\cite{Brown:1973ji}. However, neither of the contributions discussed above can break the degeneracy of the $Zb\bar{b}$ interaction owing to 
their quadratic dependence on couplings. To achieve that, one must consider the interference effects from the $\gamma$-$Z$ process (black line, labelled as ``$\gamma$-$Z$ interference" in Fig.~\ref{fig:Afb_part}; the interference between Fig.~\ref{ga_gaISR}+Fig.~\ref{ga_gaFSR} and Fig.~\ref{z_gaISR}+Fig.~\ref{z_gaFSR}), which is linearly dependent on the $Zb\bar{b}$ coupling.

\begin{figure}
    \centering
    \includegraphics[scale=0.6]{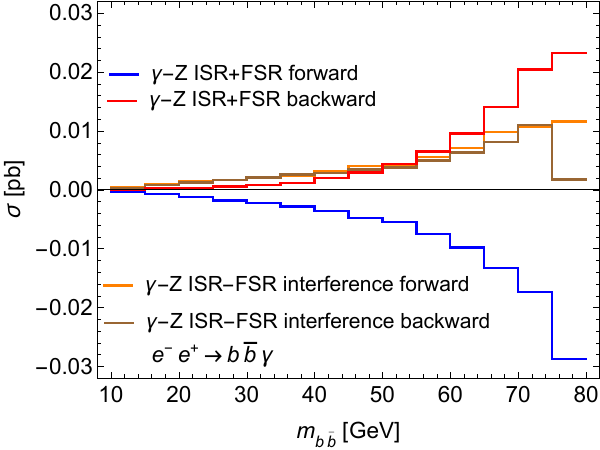}
\caption{The forward and backward cross sections of $\gamma$-$Z$ interference channel for the $e^+ e^- \to b \bar{b} \gamma$ production at the $Z$-pole in the SM.}
\label{fig:xs_za}
\end{figure}

To clarify the source of the $A_{\rm FB}^b$ in the $\gamma$-$Z$ interference process, we present the forward and backward cross sections from the ISR + FSR and the interference between the ISR and FSR in Fig.~\ref{fig:xs_za}. This asymmetry is related to the difference between the forward and backward cross sections, as a result, it is dominated by the ISR + FSR contributions themselves (blue and red lines), while the contributions from the interference between the ISR and FSR are very small (orange and brown lines). This is because the asymmetry is proportional to $g_Vg_V^e$, where $g_V^e$ is the vector component of the SM $Ze\bar{e}$ coupling, $g_V^e=-1/2+2s_W^2\simeq -0.038$, which is nearly zero, and can be understood from the charge conjugation transformation for the final states~\cite{Brown:1973ji}. As a result, the interference term becomes negligible for the $A_{\rm FB}^b $ due to the small value of $g_V^e$. 
The situation for the $e^+ e^- \to b \bar{b} g$ process is similar to the $b\bar{b}\gamma$ channel and would be simpler, as only the on-shell $Z$ production channel and the $\gamma$-$Z$ interference channel contribute to the asymmetry.

\begin{figure}
\subfigure{
    \centering
    \includegraphics[width=0.37\textwidth]{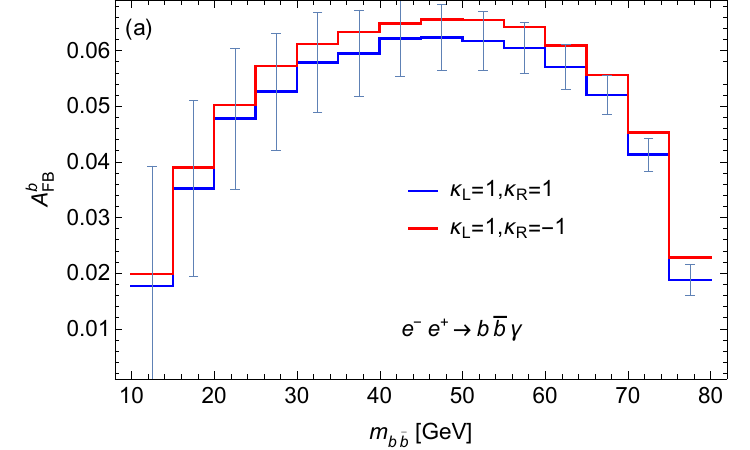}
}
\subfigure{
    \centering
    \includegraphics[width=0.37\textwidth]{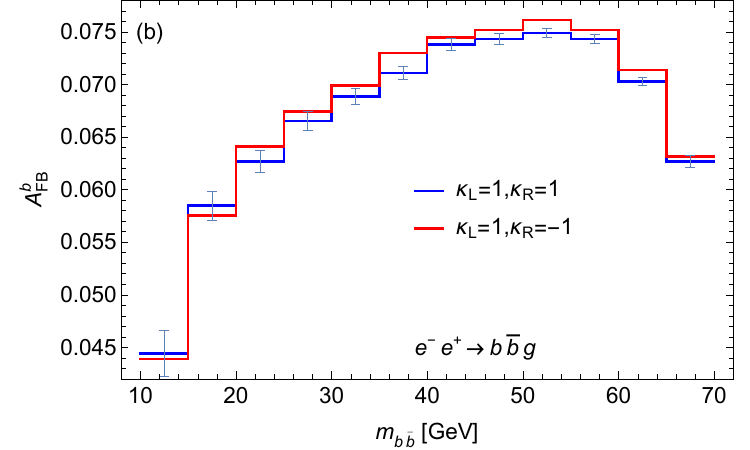}
}
\caption{The $A^{b}_\text{FB}$ distributions, with $\kappa_{L,R}= 1$ (blue) and $\kappa_L = -\kappa_R = 1$ (red), of the $e^+ e^- \to b \bar{b} \gamma$ (a) and $e^+ e^- \to b \bar{b} g$ (b) productions at the $Z$-pole.
  The error bar is the expected statistical uncertainty of measurement for each bin at the CEPC.
}
\label{fig:comp_AFB}
\end{figure}

Figure~\ref{fig:comp_AFB} presents the $A_{\rm FB}^b$ distributions from $e^+ e^- \to b \bar{b} \gamma$ (a) and $e^+ e^- \to b \bar{b} g$ (b) processes at the $Z$-pole. The blue lines represent the SM case ($\kappa_{L,R}=1$), while the red lines correspond to the wrong-sign solution for the $Zb_R\bar{b}_R$ coupling ($\kappa_L=-\kappa_R=1$). Similar to the $b\bar{b}\gamma$ channel, we require a transverse momentum of the light jet $p_T^j>20~{\rm GeV}$ for the $b\bar{b}g$ process. In the SM, the $\gamma$-$Z$ interference gives a negative contribution to $A_{\rm FB}^b$ (see Fig.~\ref{fig:Afb_part}), as a result, the wrong-sign solution will enhance this asymmetry.
To assess the possibility of distinguishing between these two parameter scenarios, we also show the expected uncertainty for each bin at the CEPC running at the $Z$-pole by rescaling the uncertainties as follows: 
\beq 
\delta (A^b_{\text{FB}})i = \sqrt{\frac{\sigma(e^+ e^- \to b \bar{b})}{\sigma(e^+ e^- \to b \bar{b} \gamma/g)i}} \delta A^{0,b}_{\text{FB}}, 
\label{eq:err}
\eeq
where $\delta A^{0,b}_{\text{FB}}=5.4\times 10^{-5}$ is the projected statistical uncertainty of $A_{\rm FB}^{0,b}$ at the CEPC~\cite{Gori:2015nqa}. It is important to note that, due to the large statistical uncertainties for each bin; see Fig.~\ref{fig:comp_AFB}, we will neglect the systematic errors in the following analysis.

\begin{figure}[tb]
\subfigure{
    \centering
    \includegraphics[scale=0.5]{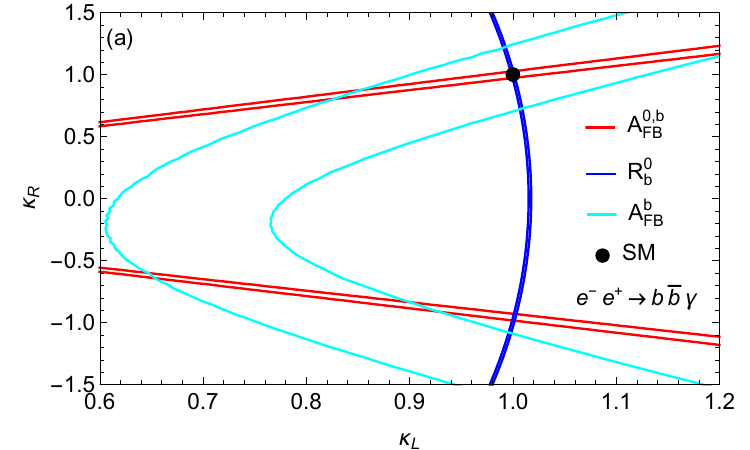}
}
\subfigure{
    \centering
    \includegraphics[scale=0.5]{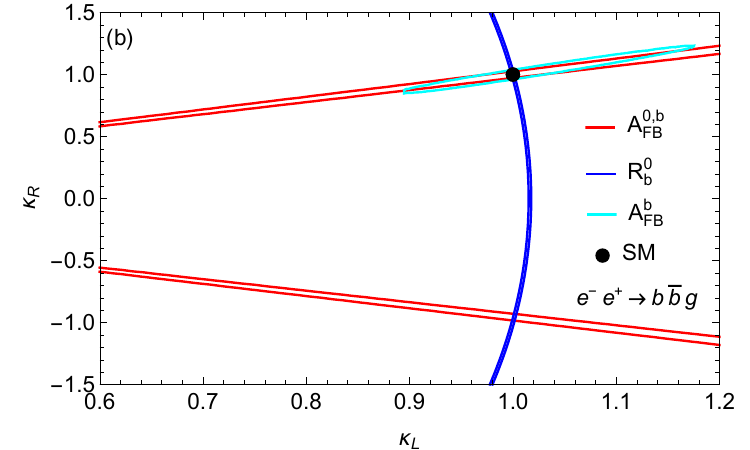}
}
\caption{The expected limits on the anomalous $Zb\bar{b}$ couplings $\kappa_L$ and $\kappa_R$ from measuring the $A_{\rm FB}^b$ distributions of the $e^+ e^- \to b \bar{b} \gamma$ (a) and $e^+ e^- \to b \bar{b} g$ (b) productions at the CEPC, running at the Z-pole, at the 95\% C.L. (cyan contours). The blue and red contours come from the expected $R^0_b$ and $A^{0,b}_\text{FB}$ measurements at the CEPC, respectively. Both the statistical and systematic errors have been included for the $R_b^0$ and $A_{\rm FB}^{0,b}$ measurements. We also assume the experimental values agree with the SM predictions. 
}
\label{fig:para_space}
\end{figure}

\section{Sensitivity at the CEPC}
In this section, we utilize the $A_{\rm FB}^b$ distributions from $e^+e^- \to b\bar{b}\gamma/g$ at the $Z$-pole to probe the $Zb\bar{b}$ coupling. By performing the pseudo experiments, we conduct a combined $\chi^2$ analysis as, 
\beq
\chi^2 = \sum_i \left[\frac{(A^b_{\text{FB}})^{\text{th}}_i-(A^b_{\text{FB}})^{\text{SM}}_i}{\delta (A^b_{\text{FB}})_i} \right]^2\,,
\eeq
where $(A_{\rm FB}^b)_i^{\rm SM}$ is the forward-backward asymmetry of the SM (i.e., $\kappa_{L,R}=1$) for the $i$-th bin. The statistical uncertainty $\delta (A_{\rm FB}^b)_i$ is estimated from the $A_{\rm FB}^{0,b}$ measurement at the $Z$-pole, see Eq.~\eqref{eq:err}. For simplicity, we have assumed that the experimental values of the $(A_{\rm FB}^b)_i$ agree with the SM predictions.

In Fig.~\ref{fig:para_space}, we show the expected limits (cyan contours) on the $Zb\bar{b}$ coupling from the $e^+e^-\to b\bar{b}\gamma$ (a) and $e^+e^-\to b\bar{b}g$ (b) at the 95\% confidence level (C.L.) at the CEPC running at the $Z$-pole. To uniquely determine the $Zb\bar{b}$ coupling, we plot the expected limits from the $A_{\rm FB}^{0,b}$ (red contours) and $R_b^0$ (blue contours) at the CEPC in the same figure. Here, we focus on the parameter space with $\kappa_L>0$ which has been verified by the off-$Z$ pole $A_{\rm FB}^b$ measurements at the LEP. It clearly shows that the $Zb\bar{b}$ coupling could be determined uniquely with the measurements at the $Z$-pole alone at the future lepton colliders.

\section{Conclusions}
In this paper, we propose a novel approach to probe the $Zb\bar{b}$ coupling by studying the forward-backward asymmetry distributions as a function of $m_{b\bar{b}}$ in the center-of-mass frame of the $b\bar{b}$ system from $e^+ e^- \to b \bar{b} \gamma$ and/or $e^+ e^- \to b \bar{b} g$ processes at the CEPC running at the $Z$-pole. We demonstrate that the $\gamma$-$Z$ interference effects play a crucial role in shaping the $A_{\rm FB}^b$ distributions and show these effects are linearly sensitive to the $Zb\bar{b}$ coupling parameters, $\kappa_{L,R}$, and provide complementary information to the traditional measurements of $R^0_b$ and $A^{0,b}_\text{FB}$ at the $Z$-pole, which primarily probe the quadratic terms of $\kappa_L$ and $\kappa_R$.
The forward-backward asymmetry distribution measurements from these processes offers a straightforward and effective means of resolving the sign degeneracy  of $\kappa_R$ observed in precision electroweak data from LEP and SLC experiments. This approach does not require any special optimization, making it a promising avenue for studying the $Zb\bar{b}$ coupling at the future lepton colliders.

\vspace{3mm}
\noindent{\bf Acknowledgments.}
Bin Yan is supported by the IHEP under Contract No. E25153U1. 
C.-P. Yuan is supported by the U.S. National Science Foundation under Grant No. PHY-2013791 and the support from the Wu-Ki Tung endowed chair in particle physics. Shu-Run Yuan is supported in part by the National Science Foundation of China under Grants No.11725520, No.11675002 and No.12235001.

\bibliographystyle{apsrev}
\bibliography{reference}

\end{document}